\documentclass[aps,pra,showpacs,amsfonts,amssymb,amsmath,twocolumn]{revtex4} 
\usepackage{graphicx}
\usepackage{color}
\usepackage{epsfig}
\usepackage{times}

\renewcommand{\vec}{\mathbf}
\newcommand{\leftexp}[2]{{\vphantom{#2}}^{#1}{#2}}

\begin{document}

\title{Rotational master equation for cold laser-driven molecules}

\author{A.~Adelsw\"ard}

\affiliation{Emmy--Noether Nachwuchsgruppe ``Kollektive Quantenmessung
  und R\"uckkopplung an Atomen und Molek\"ulen'', Fachbereich Physik,
  Universit\"at Rostock, Universit\"atsplatz 3, D-18051 Rostock,
  Germany}

\author{S.~Wallentowitz}

\affiliation{Emmy--Noether Nachwuchsgruppe ``Kollektive Quantenmessung
  und R\"uckkopplung an Atomen und Molek\"ulen'', Fachbereich Physik,
  Universit\"at Rostock, Universit\"atsplatz 3, D-18051 Rostock,
  Germany}

\email{sascha.wallentowitz@physik.uni-rostock.de}

\author{W.~Vogel}

\affiliation{Arbeitsgruppe Quantenoptik, Fachbereich Physik,
  Universit\"at Rostock, Universit\"atsplatz 3, D-18051 Rostock,
  Germany}

\date{December 13, 2002}

\begin{abstract}
  The equations of motion for the molecular rotation are derived for
  vibrationally cold dimers that are polarized by off-resonant laser
  light. It is shown that, by eliminating electronic and vibrational
  degrees of freedom, a quantum master equation for the reduced
  rotational density operator can be obtained. The coherent rotational
  dynamics is caused by stimulated Raman transitions, whereas
  spontaneous Raman transitions lead to decoherence in the motion of
  the quantized angular momentum.  As an example the molecular
  dynamics for the optical Kerr effect is chosen, revealing
  decoherence and heating of the molecular rotation.
\end{abstract}

\pacs{33.80.-b, 03.65.Yz, 42.50.Ct}


\maketitle

\section{Introduction} \label{sec:1}
Nonlinear optics, as commonly understood, describes the effects of
nonlinear media on the optical field via susceptibilities, that are
derived by a perturbative approach~\cite{nonlin-optics}. It is assumed
in this context, that the medium can be treated as being stationary,
which is well justified for thermalized samples. The medium may be
specifically composed of atoms, molecules or may be implemented as a
solid-state device, such as a semiconductor. The change of the quantum
state of that physical system during the light--matter interaction is
however largely neglected in the nonlinear optics literature,
naturally since the focus lies on the properties of the light. Clearly
the treated problem is a coupled one, where a description by
susceptibilities is the outcome of a decoupling of medium and light
degrees of freedom that is only justified by certain restrictive
assumptions on the properties of the medium. A quite opposite
viewpoint is to consider the dynamics of the atomic systems under the
influence of the applied laser field~\cite{qo-book}. In the latter
case the scattered radiation is of minor importance, only its effect
on the atomic dynamics by absorption and subsequent stimulated and
spontaneous emissions of photons is taken into account.  Thus the
ancillary medium of nonlinear optics becomes the central issue.

Focusing on the atomic or molecular dynamics however requires a
treatment of physical conditions that are more general than those
usually considered in nonlinear optics.  For instance, if significant
dynamics emerge, non-stationary initial quantum states and/or
sufficiently strong coherent laser drives prevail. For atoms such
situations have been reached in several implementations, the majority
of them by providing low temperatures and atom-light coupling
strengths comparable or larger than spontaneous decay rates.
Prominent examples are atoms in high-Q
microwave~\cite{walther-haroche,haroche-qnd,walther2} or
optical~\cite{kimble-rempe} cavities and single ions in radio-frequency
traps~\cite{ion-cooling,ion-coupling}.

Recent advances in trapping and cooling of
atomic~\cite{atom-cooling,alkali-BEC,hydrogen-BEC,atom-laser,atom-microstructure}
and
molecular~\cite{pendulum,mol-trap-dipole,mol-trap-magn,mol-photoass,mol-photoass-bec}
gases, progressively give access to the quantum-statistical properties
and quantum dynamics of the manipulated gases. The question therefore
arises as to how the quantum state of molecular gases, for example,
can be manipulated by nonlinear optical interactions or how some of
its properties can be revealed from the nonlinear optical spectra. For
this purpose it is necessary to go beyond standard calculations of
nonlinear optics, in order to treat not only the optical-field
dynamics, but to focus on the dynamics of the molecular system.

A first step towards such considerations is the study of the
rotational dynamics of molecules that interact with off-resonant laser
fields.  Since dimers have no permanent dipole moment, they are
infrared inactive and thus the rotational dynamics is expected to be
to large extent undamped. However, the polarizability of the molecule
enables two-photon transitions and it will be shown here that these
may in fact lead to decoherence in the rotational dynamics.

Another aspect makes this issue particularly interesting. Consider
ultracold molecules in a trap that is based on off-resonant optical
fields, such as a dipole trap~\cite{mol-trap-dipole}. Assuming the
molecules are cool enough to form a degenerate Bose gas, the lifetime
of this peculiar quantum state is of major interest. As we will see
here, off-resonant laser fields lead to a decoherence in the
rotational quantum state, meaning that after some time the initial
rotational quantum state evolves into a statistical mixture of
differing angular-momentum states. In that case one may expect that
also the degenerate condensate state of molecules will be destroyed.

In this paper we derive the basic equations of motion that describe
the molecular rotational dynamics. We will focus on homonuclear,
diatomic molecules interacting with both weak, off-resonant laser
fields and electromagnetic vacuum modes.  In such a configuration,
besides the typical elastic Rayleigh scattering, stimulated and
spontaneous Raman scattering processes occur that affect the
molecular, ro-vibrational quantum state and may lead to decoherence.
We derive a master equation that includes these processes for studying
rotational decoherence, which may play an important role in the
context of recent experiments with ultracold
molecules~\cite{mol-photoass-bec}. The theory is applied to the optical
Kerr effect, for which we solve the master equation by means of
quantum trajectories.

The paper is structured as follows: In Sec.~\ref{sec:2} the physical
situation under study is explained and the basic equations for the
molecule-light interaction are given. Sec.~\ref{sec:3} is devoted to
the adiabatic elimination of electronically excited states. In
Sec.~\ref{sec:4} the trace over the vibrational degree of freedom is
performed, leading to the reduced rotational master equation.  It is
solved in Sec.~\ref{sec:6} for the case of an optical Kerr interaction
and the damping and decoherence effects in the rotational degree of
freedom are illustrated. Finally in Sec.~\ref{sec:7} a summary and
conclusions are given.

\section{Molecule-light interaction} \label{sec:2}
\begin{figure}
  \begin{center}
    \epsfig{file=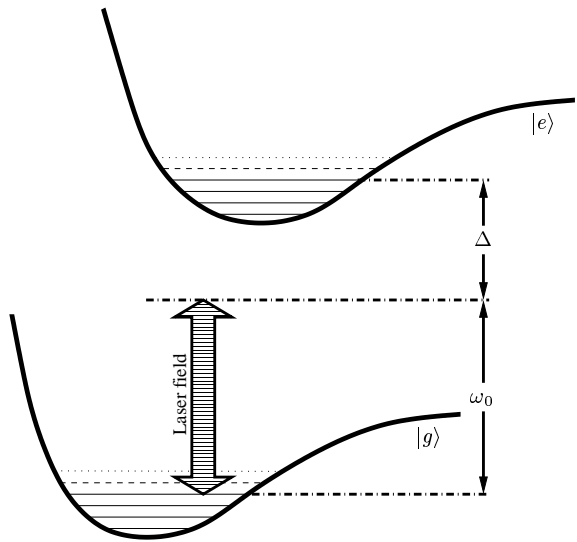,scale=1}
    \caption{Off-resonant laser excitation scheme of the
      molecule: $|g\rangle$ and $|e\rangle$ are electronic ground
      and excited states with respective inter-nuclear potentials,
      $\Delta$ is the detuning of the laser frequency $\omega_0$ to
      the bare electronic transition frequency.}
    \label{fig:level-scheme}
  \end{center}
\end{figure}

We consider a cold, homonuclear molecule interacting with a light
pulse far detuned from possible resonances connected with
electronic-state transitions of the type $\leftexp{1}{\Sigma}
\!\leftrightarrow\!  \leftexp{1}\Sigma$. More precisely, the molecule
is assumed to be initially in its electronic ground-state potential
with a given distribution of populations among the ro-vibrational
states. The laser field is then considered to be off-resonant, if all
allowed ro-vibrational transitions from this distribution to excited
electronic states are off-resonant, cf.~Fig.~\ref{fig:level-scheme}.
In this way two-photon Raman transitions dominate the dynamics and we
consider in the following stimulated as well as spontaneous Raman
processes.

Since electronic, vibrational and rotational transition frequencies
are different from each other by orders of magnitude,
\begin{equation}
  \label{eq:timescales}
  \omega_{eg} \gg \omega_\nu \gg B ,
\end{equation}
the dynamics reveals a separation of timescales.  In
Eq.~(\ref{eq:timescales}) $\omega_{eg}$ is the bare transition
frequency between the ground and nearest (dipole-allowed) excited
electronic potential minima. The frequencies $\omega_\nu$ and $B$ are
the typical vibrational frequency and the rotational constant,
respectively. To avoid vibrational Raman transitions the spectral
width $\Delta \omega$ of the laser pulse obeys then the requirement
\begin{equation}
  \label{eq:long-pulse}
  \Delta \omega \ll \omega_\nu ,
\end{equation}
and is typically of the order of rotational frequencies: $\Delta\omega
\!\sim\! B$. Typically we may think of the laser field as being
bichromatic for addressing particular rotational transitions, though
also more general spectra are possible. Note, that
condition~(\ref{eq:long-pulse}) is contrary to the situation in
vibrational wavepacket creation by ultrashort laser
pulses~\cite{fs-wavepacket}.

We assume the molecule being located at a fixed position and omit its
center-of-mass coordinate. Thus with respect to its center of mass a
point-like molecule is considered.  However, our derivations can in
principle be extended to include the quantum mechanical center-of-mass
motion of the molecule. This extension would apply for describing
molecular beam deflection~\cite{deflection-stapelfeldt,
  deflection-sakai} and would extend the purely coherent
treatment~\cite{deflection-domokos} by including spontaneous
processes.

We start with the general dynamical equations of a molecule
interacting with a generic laser field and spontaneously emitting
photons into the vacuum electromagnetic reservoir. The master equation
for the density operator of electronic, vibrational and rotational
degrees of freedom $\hat{\varrho}$ can be written as
\begin{equation}
  \label{eq:master}
  \dot{\hat{\varrho}} = - \frac{i}{\hbar} [ \hat{H}, \hat{\varrho} ] + 
  \hat{\cal L}_{\rm s} \, \hat{\varrho}  \; .
\end{equation}
The time-dependent Hamiltonian reads
\begin{equation}
  \label{eq:ham-total}
  \hat{H}(t) = \hat{H}_{\rm M} + \hat{H}_{\rm ML}(t) \; ,
\end{equation}
where $\hat{H}_{\rm M}$ is the free Hamiltonian of the molecule and
$\hat{H}_{\rm ML}(t)$ describes the interaction of the molecule with
the applied classical laser field.

The Hamiltonian of the molecule can be written in the form
\begin{equation}
  \label{eq:ham-mol}
  \hat{H}_{\rm M} = \sum_g \hbar \omega_g |g\rangle \langle g| +
  \sum_e \hbar \omega_e |e\rangle \langle e| \; ,
\end{equation}
where $|g\rangle$ and $|e\rangle$ are short-hand notations for
ro-vibrational energy eigenstates in the ground and excited electronic
potentials, respectively,
\begin{equation}
  \label{eq:9}
  |g\rangle = |g,\nu_g,j_g,m_g\rangle, \qquad |e\rangle =
   |e,\nu_e,j_e,m_e\rangle \; .
\end{equation}

Our derivation applies to general molecular energy spectra. Only the
ro-vibrational coupling needs to be neglected here. This coupling has
been shown to lead to a decohering effect in the vibrational
motion~\cite{brif-decoherence,wal-decoherence} and thus can be
expected to produce decoherence also in the rotational degree of
freedom. It appears, however, on a much larger timescale compared with
the rotational frequencies and can be neglected if the molecule is and
stays vibrationally cold during the interaction.  Spontaneous Raman
processes lead to a diffusion in the vibrational state, which may heat
the vibrational degree of freedom. The diffusion rate, however, is
small enough for being neglected.  In simplest approximation, for
example, assuming a harmonic inter-nuclear vibration and taking into
account only lowest-order rotational contributions, the
eigenfrequencies could be given as
\begin{eqnarray}
  \label{eq:mol-frequencies}
  \omega_g & = & \omega_{\nu,g} ( \nu_g \!+\!
  {\textstyle\frac{1}{2}} ) + B_g \, j_g (j_g \!+\! 1) \; , \\
  \label{eq:mol-frequencies2}
  \omega_e & = & \omega_{eg} + \omega_{\nu,e} ( \nu_e \!+\!
  {\textstyle\frac{1}{2}} ) + B_e \, j_e (j_e \!+\! 1) \; .
\end{eqnarray}
Here $\omega_{\nu,g/e}$ are the vibrational frequencies in the ground
and excited electronic potential, and $B_{g/e}$ are the rotational
constants in these electronic states. 

Turning back to Eq.~(\ref{eq:ham-total}) the interaction of the
molecule with the applied laser field $\hat{\vec{E}}(t)$ reads 
\begin{equation}
  \label{eq:ham-int}
  \hat{H}_{\rm int}(t) = - \hat{\vec{d}} \!\cdot\! \vec{E}(t) \; ,
\end{equation}
where $\hat{\vec{d}}$ is the dipole operator of the single molecule.
The laser field can be decomposed into slowly varying positive and
negative frequency components $\vec{E}^{(\pm)}(t)$ by using the ansatz
\begin{equation}
  \label{eq:e-field}
  \vec{E}(t) = \vec{E}^{(+)}(t) \, e^{-i\omega_0 t} + \vec{E}^{(-)}(t)
  \, e^{i\omega_0 t} ,
\end{equation}
where $\omega_0$ is the mid frequency of the laser pulse. Then the
vector amplitudes $\vec{E}^{(\pm)}(t)$ oscillate only at a maximum
frequency given by the spectral width $\Delta \omega$, which is of the
order of rotational frequencies.

The incoherent part of the master equation~(\ref{eq:master}) describes
spontaneous emissions of photons into the vacuum electromagnetic
reservoir and is given by the Liouville operator $\hat{\cal L}_{\rm
  s}$. For the molecular energy levels being degenerate with respect
to the rotational quantum number $m$ the action of $\hat{\cal L}_{\rm
  s}$ on the density operator reads~\cite{cohen-tannoudji}
\begin{eqnarray}
  \label{eq:L-spon}
  \lefteqn{\hat{\mathcal L}_{\rm s} \, \hat{\varrho} = \sum_{i=1}^3
    \sum_{ab} \sum_{a'b'} |a\rangle \langle b| \, 
    \delta(\omega_{ab} \!-\! \omega_{a'b'}) } & & \nonumber \\ 
  & & \times \Big[ \langle a|\hat{\gamma}_i|a' \rangle
  \langle a' | \hat{\varrho} | b' \rangle \langle b' | 
  \hat{\gamma}_i^\dagger | b \rangle - {\textstyle\frac{1}{2}} 
  \langle a | \hat{\gamma}_i^\dagger  \hat{\gamma}_i | a'
  \rangle  \langle a' | \hat{\varrho} | b' \rangle  \langle b' | b
  \rangle \nonumber \\ 
  & & \quad - \, {\textstyle\frac{1}{2}}
  \langle a | a' \rangle \langle a' | \hat{\varrho} 
  | b' \rangle \langle b' | \hat{\gamma}_i^\dagger \hat{\gamma}_i 
  | b \rangle \Big] \; .
\end{eqnarray}
The function $\delta(\omega \!-\! \omega')$ is a quasi
delta function that ensures energy conservation $\omega \!\approx\!
\omega'$ for optical frequencies,
\begin{equation}
  \label{eq:13}
  \delta(\omega) = \left\{ \begin{array}{cl} 
      1 & \mbox{for $|\omega| \!\ll\! \omega_0$} , \\
      0 & \mbox{else} .
    \end{array} \right. \; 
\end{equation}
It selects resonant terms resulting from the performed secular
approximation. Furthermore, the notation 
\begin{equation}
  \omega_{ab} =  \omega_a - \omega_b ,
\end{equation}
denotes transition frequencies between the ro-vibronic molecular
states $|a\rangle$ and $|b\rangle$ ($a,b \!=\! g,e$).

The operators $\hat{\gamma}_i$ reflect the spontaneous emission of a
photon due to molecular dipole moment pointing into direction $i$.
That is, $i$ labels the polarization of the spontaneously emitted
photon. These operators are defined as
\begin{equation}
  \label{eq:def-gamma-i}
  \hat{\gamma}_i = \kappa \, \hat{d}_i \qquad 
  (i = 1,2,3) ,
\end{equation}
where $\hat{d}_i$ is a vector component of the dipole moment and
\begin{equation}
  \label{eq:def-kappa}
  \kappa = \sqrt{\frac{\omega_{eg}^3}{3\pi c^3 \epsilon_0 \hbar}} \; .
\end{equation}
For Eq.~(\ref{eq:def-kappa}) tiny modifications due to ro-vibrational
frequencies have been neglected, since they are much smaller than the
bare (optical) electronic transition frequency $\omega_{eg}$.

Note that the Liouville operator~(\ref{eq:L-spon}) looks different
from the usual cases considered in standard quantum-optical systems,
such as two-level atoms. This is only due to the energetic degeneracy
with respect to the rotational quantum number $m$.  To properly
account for this degeneracy we have to allow for subtle coherent
effects: Starting from a ro-vibronic state $|e,\nu_e,j_e,m_e\rangle$,
the spontaneous emission of a photon of frequency $\omega$ may leave
the molecule in a coherent superposition of ro-vibronic quantum states
$|g,\nu_g,j_g,m_g\rangle$ with a unique value for $\nu_g$ and $j_g$
but multiple values $m\!\in\![-j_g, j_g]$ according to the selection
rules.

Given the dynamical equation~(\ref{eq:master}) of the molecule
interacting with the off-resonant laser pulse, in the following
sections we will derive a master equation for the reduced rotational
density operator $\hat{\sigma}$, of the form
\begin{equation}
  \dot{\hat{\sigma}} = - \frac{i}{\hbar} [ \hat{H}_{\rm M} \!+\! 
  \hat{H}_{\rm R}, \hat{\sigma} ] + \hat{\mathcal L}_{\rm sR} \, 
  \hat{\sigma} \; .
\end{equation}
The effective two-photon Hamiltonian $\hat{H}_{\rm R}$ will contain
the coherent drive due to optically stimulated rotational Raman
transitions. It contains the polarizability tensor of the molecule and
the classical laser field.  The incoherent effects due to spontaneous
Raman scattering will be described by the Liouvillean operator
$\hat{\mathcal L}_{\rm sR}(t)$.

\section{Elimination of excited electronic states} \label{sec:3}
For the interaction with an off-resonant laser pulse the molecule is
hardly driven from the ground electronic potential surface into the
excited one. This applies, given that not too high vibrational levels
are populated, which is the case for cold molecules as considered
here.  Thus the electronically excited states $|e\rangle \!=\!
|e,\nu_e,j_e,m_e\rangle$ can be adiabatically eliminated to
second-order in the molecule-light coupling. For this purpose we start
with the coupled equations of motion of density-matrix elements,
derived from the master equation~(\ref{eq:master})
\begin{eqnarray}
  \label{eq:gg'}
  \dot{\varrho}_{gg'} & = & -i \omega_{gg'} \varrho_{gg'} + i 
  \sum_{e} \left[ \Omega^\ast_{eg}   
    \varrho_{eg'} \, e^{ i \omega_0 t } \!-\! \varrho_{ge}  \Omega_{eg'}
    \, e^{ - i\omega_0 t} \right] 
  \nonumber \\
  & & + \sum_{i} \sum_{ee'} \gamma_{i,ge} \, \varrho_{ee'} \, 
  \gamma^{\ast}_{i,g'e'} \; , \\
  \label{eq:ee'}
  \dot{\varrho}_{ee'} & = & - i\omega_{ee'} \varrho_{ee'} + i
  \sum_g \left[
    \Omega_{eg} \varrho_{ge'}  \, e^{-i\omega_0 t} \!-\! 
    \varrho_{eg} \Omega^\ast_{e'g} \, e^{i\omega_0 t} \right] 
  \nonumber \\
  & & - \, \Gamma \, \varrho_{ee'} \; , \\
  \label{eq:eg}
  \dot{\varrho}_{eg} & = & - i \omega_{eg} \varrho_{eg} + i \sum_{g'}
  \Omega_{eg'} \varrho_{g'g}  \, e^{-i\omega_0 t}
  \nonumber \\
  & & - i \sum_{e'} \varrho_{ee'} \Omega_{e'g} \, e^{-i\omega_0 t} 
  - {\textstyle\frac{1}{2}} \Gamma \, \varrho_{eg} \; , 
\end{eqnarray}
and $\dot{\varrho}_{ge} \! = \! \dot{\varrho}_{eg}^\ast$.  Here
$\varrho_{ab}$ are the matrix elements of the density operator in
energy eigenstates
\begin{eqnarray}
  \label{eq:def-dens-mat}
  \varrho_{ab} & = & \langle a | \, \hat{\varrho} \, | b \rangle 
  \\ \nonumber
  & = & \langle a,
  \nu_a, j_a, m_a | \, \hat{\varrho} \, | b, \nu_b, j_b, m_b \rangle  
  \qquad  (a,b = g,e) \; ,
\end{eqnarray}
and $\gamma_{i,ge}$ are the matrix elements of the
spontaneous-emission operators $\hat{\gamma}_i$:
\begin{equation}
  \gamma_{i,ge} = \langle g | \hat{\gamma}_i | e\rangle \; .
\end{equation} 
Furthermore, in Eqs~(\ref{eq:ee'}) and (\ref{eq:eg}) the total decay
rate $\Gamma$ is defined as
\begin{eqnarray}
  \label{eq:Gamma}
  \Gamma = \sum_i \sum_g \gamma_{i,ge}^\ast \gamma_{i,ge}
  = \kappa^2
  d^2 = \frac{d^2 \omega_{eg}^3}{3 \pi c^3 \epsilon_0 \hbar} \; ,
\end{eqnarray}
with $d$ being the matrix element of the dipole moment of the
considered electronic transition (cf. App.~\ref{sec:app1}).  Finally,
the time-dependent single-photon Rabi frequencies $\Omega_{eg}(t)$ are
defined as
\begin{equation}
  \label{eq:rabi}
  \Omega_{eg}(t)  = \frac{1}{\hbar} \, \langle e|
  \hat{\vec{d}} | g \rangle \!\cdot\! \vec{E}^{(+)}(t) \; .
\end{equation}
They are slowly varying at frequencies given by the spectral width
$\Delta\omega$ of the laser pulse.

We now formally integrate the equation of motion for the electronic
coherences, Eq.~(\ref{eq:eg}), from an initial time $t'$ to a later
time $t$, that obey the condition
\begin{equation}
  \label{eq:adiab-cond}
  \frac{1}{\Delta} \ll t \!-\! t' \ll \frac{1}{\Omega}, \frac{1}{\Gamma} ,
\end{equation}
and insert the obtained solutions into the equation of motion for the
ground-state populations, Eq.~(\ref{eq:gg'}). We consistently keep
only terms up to second-order in the molecule-light interaction.
Defining the slowly varying density-matrix elements
$\tilde{\varrho}_{ab}(t)$,
\begin{equation}
  \label{eq:slow-dens-mat}
  \varrho_{ab}(t) = \tilde{\varrho}_{ab}(t) \, e^{-i\omega_{ab} t} , 
\end{equation}
the approximation $\tilde{\varrho}_{ab}(t \!-\! \tau) \!\approx\!
\tilde{\varrho}_{ab}(t)$ can be performed for times $\tau$ within the
integration interval $t \!-\! t'$. From this procedure the equation of
motion for the electronic ground-state manifold is obtained,
\begin{widetext}
\begin{eqnarray}
  \dot{\varrho}_{gg'} & = & -i\omega_{gg'} \varrho_{gg'} 
  + \sum_i \sum_{ee'} \gamma_{i,ge}  \varrho_{ee'}  \gamma^{\ast}_{i,g'e'} 
  + \Bigg\{ \sum_{e''} \Omega^\ast_{e''g}(t) \bigg[ 
  \sum_{e'} \tilde{\rho}_{e''e'}(t) \, e^{-i\omega_{e''e'}t} 
  \int_0^{t-t'} \!\!\! d\tau \, \Omega_{e'g'}(t \!-\! \tau) \, 
  e^{i(\omega_0 - \omega_{e'g'}) \tau - \frac{1}{2} \Gamma \tau}  
  \nonumber \\
  & & \qquad\qquad - \, \sum_{g''} \tilde{\rho}_{g''g'}(t) \, 
  e^{-i\omega_{g''g'} t} 
  \int_0^{t-t'} \!\!\! d\tau \, \Omega_{e''g''}(t \!-\! \tau) 
  \, e^{i(\omega_0 - \omega_{e''g''}) \tau - \frac{1}{2} \Gamma \tau} 
  \bigg]     + ( \mbox{c.c. and $g \!\leftrightarrow\! g'$} )
  \Bigg\} \; .
\end{eqnarray} 
\end{widetext}

The time dependent Rabi frequencies $\Omega_{eg}(t)$ oscillate much
slower than the molecular vibrational frequencies since $\Delta\omega
\!\ll\!  \omega_{\nu,g/e}$. They are slowly varying as
compared to the oscillating exponentials in the integrals, that on
average oscillate with the bare laser detuning defined by
\begin{equation}
  \Delta = \omega_0 -  \omega_{eg} \; .
\end{equation}
Thus $\Omega_{eg}(t \!-\! \tau) \!\approx\! \Omega_{eg}(t)$ for the
time integrations and therefore the Rabi frequencies can be taken
outside the integral. After performing the remaining integrations and
neglecting all remaining terms that oscillate with frequencies of the
order of the detuning $\Delta$ or the vibrational frequency
$\omega_{\nu,g/e}$, we obtain
\begin{widetext}
\begin{eqnarray}
  \label{eq:gg'-eg-2}
  \dot{\varrho}_{gg'} & = &  -i\omega_{gg'} \varrho_{gg'}
  + \sum_{e g''} \left[ \varrho_{gg''} \delta_{\nu_g \nu_{g''}} 
  \, \frac{\Omega_{eg''}^\ast(t) \, \Omega_{eg'}(t)} 
  {\Delta^2} \, (-i\Delta \!-\! {\textstyle\frac{1}{2}} 
  \Gamma) + (\mbox{c.c. and $g
  \!\leftrightarrow\! g'$}) \right] 
  \nonumber \\
  & &  + \sum_{ee'} \varrho_{ee'} \left[ \sum_i   
  \gamma_{i,ge} \, \gamma^{\ast}_{i,g'e'}
  +  \delta_{\nu_e \nu_{e'}} \, 
  \frac{\Omega_{eg}^\ast(t) \, \Omega_{e'g'}(t)}{\Delta^2} \, \Gamma 
  \right] .
\end{eqnarray}
\end{widetext}
Here $\nu_g$ and $\nu_e$ are the numbers of vibrational quanta in the
ground and excited electronic potentials, respectively, and the
Kronecker delta $\delta_{\nu_g \nu_{g''}}$ and $\delta_{\nu_e
  \nu_{e'}}$ result as resonances from the rotating-wave
approximation. 
This rotating-wave approximation with respect to vibrational frequencies holds
here because we will consider from now on only density-matrix elements
diagonal in the vibrational quantum numbers (in Eq.~(\ref{eq:gg'-eg-2}) $\nu_g
\!=\! \nu_{g'}$). Those matrix elements do not freely oscillate with
vibrational frequencies and since also the electric-field spectrum is
narrower, they evolve on a slow timescale given by the Raman drive. 
Moreover, we have neglected small variations of the
detuning due to ro-vibrational frequencies, since in the far
off-resonant case considered here we may safely assume the condition
\begin{equation}
  \Delta \gg \omega_{\nu,g/e}, B_{g/e} \; .
\end{equation}

Considering a laser pulse that is well off-resonant to the electronic
transition means also that the bare detuning is larger than the
natural linewidth of the dipole transition by orders of magnitude,
i.e.,
\begin{equation}
  \Delta \gg \Gamma \; .
\end{equation}
Since for the case of a homonuclear molecule typically $B_{g/e}
\!\gg\! \Gamma$, for the matter of consistency the original
denominators in Eq.~(\ref{eq:gg'-eg-2}) have been further approximated
by use of $[\Delta^2 \!+\! (\Gamma/2)^2]^{-1} \!\approx\!
\Delta^{-2}$.

Equation~(\ref{eq:gg'-eg-2}) is still coupled to the electronically
excited states $\varrho_{ee'}$.  Thus, the manifold of density-matrix
elements of electronically excited states is required. Performing the
same procedure as for arriving at Eq.~(\ref{eq:gg'-eg-2}) for the
electronically excited states, we obtain from Eqs~(\ref{eq:ee'}) and
(\ref{eq:eg}),
\begin{equation}
  \label{eq:ee'-eg-2}
  \varrho_{ee'} = \sum_{gg'}  \varrho_{gg'} 
  \delta_{\nu_g \nu_{g'}} \, \frac{\Omega_{eg}(t) \, 
    \Omega_{e'g'}^\ast(t)}{\Delta^2} \; . 
\end{equation}
Again this result is accurate to second-order in the molecule-light
interaction
and is valid for matrix elements diagonal in the vibrational degree of freedom
($\nu_e \!=\! \nu_{e'}$).
This term represents the excitation of the excited state from the electronic
ground-states by absorption of a laser photon.

Inserting (\ref{eq:ee'-eg-2}) into (\ref{eq:gg'-eg-2}) and omitting
terms of order higher than second in the molecule-light coupling, we
arrive at the master equation for the consistently decoupled
electronic ground-state manifold 
($\nu_g \!=\! \nu_{g'}$)
\begin{widetext}
\begin{eqnarray}
  \label{eq:gg'-eg-ee'}
   \dot{\varrho}_{gg'} =  -i\omega_{gg'} \varrho_{gg'}
  + \sum_{e g''} \left[ \varrho_{gg''} \delta_{\nu_g \nu_{g''}} 
  \, \frac{\Omega_{eg''}^\ast(t) \, \Omega_{eg'}(t)} 
  {\Delta^2} \, (-i\Delta \!-\! \Gamma/2) + (\mbox{c.c. and $g
  \!\leftrightarrow\! g'$}) \right]
  \nonumber \\
   \quad + \, \sum_i  \sum_{ee'} 
   \sum_{g''g'''} \varrho_{g''g'''} \, \delta_{\nu_{g''} \nu_{g'''}} \,
   \frac{\Omega_{eg''}(t) \, 
    \Omega_{e'g'''}^\ast(t)}{\Delta^2}
  \, \gamma_{i,ge} \, \gamma^{\ast}_{i,g'e'} \; .
\end{eqnarray}
\end{widetext}
This equation of motion describes the effects of spontaneous and
stimulated Raman scattering processes on the dynamics of the
ro-vibrational quantum state of the molecule. It represents a
consistent approximation to second-order in the molecule-light
coupling and preserves the trace of the density operator of the
ground-state manifold:
\begin{equation}
  \label{eq:11}
  \sum_g \dot{\varrho}_{gg} = \sum_{\nu_g,j_g,m_g} \langle g, \nu_g,
  j_g, m_g | \dot{\hat{\varrho}} | g, \nu_g, j_g, m_g \rangle
  = 0 \; .
\end{equation}

Note, that the eigenfrequencies $\omega_g$, that appear in
Eq.~(\ref{eq:gg'-eg-ee'}) as transition frequencies $\omega_{gg'}
\!=\! \omega_g \!-\! \omega_{g'}$, only depend on the quantum numbers
$\nu_g$ and $j_g$, i.e.,
\begin{equation} 
  \omega_g \to \omega_{\nu_gj_g} ,
\end{equation} 
which is due to the energetic degeneracy with respect to the
rotational quantum number $m_g$.

\section{Elimination of inter-nuclear vibration} \label{sec:4}
Now the vibrational degree of freedom has to be eliminated to obtain
the reduced dynamics of the molecular rotation. We intend to obtain an
equation of motion for the reduced rotational density operator
$\hat{\sigma}$, with matrix elements
\begin{equation}
  \label{eq:12}
  \sigma_{jm,j'm'} = \langle jm| \hat{\sigma} |j'm'\rangle \; ,
\end{equation}
by tracing the electronic ground-state density matrix $\varrho_{gg'}$
over the vibrational degree of freedom,
\begin{equation}
  \sigma_{jm,j'm'}  
  = \sum_{\nu_g=0}^\infty \langle g,\nu_g,j,m |\hat{\varrho}
  |g,\nu_g,j',m'\rangle \; .
\end{equation}
For notational convenience we have omitted here the
electronic-ground-state subscript $g$ for the rotational quantum
numbers: $j_g, m_g \!\to\! j,m$.

In Franck--Condon approximation the dipole operator is independent of
the inter-nuclear distance and therefore does not act in the Hilbert
space of the vibrational degree of freedom. Then the Franck--Condon
factors, being overlap integrals of vibrational wavefunctions in the
ground and excited electronic potentials
$\langle \nu_g | \nu_e \rangle$,
will naturally emerge. 
Whereas, due to the narrow spectrum of the laser pulse, the stimulated Raman
process only drives rotational transitions, spontaneous Raman scattering is
accompanied by all possible vibrational Franck--Condon transitions.
However, performing in Eq.~(\ref{eq:gg'-eg-ee'}) the trace over the
vibrational quantum number $\nu_g$ ($\nu_{g'} \!=\! \nu_g$), it comes out that
completeness relations of vibrational states can be used, so that combinations
of Franck--Condon factors are summed over to result as unity. Thus it is
possible to obtain an equation of motion for the rotation alone, that is no
longer coupled to the vibrational dynamics.

After having performed the sum over the number of vibrational quanta,
the resulting master equation for the reduced rotational density
matrix can be written in the following Lindblad form
\begin{eqnarray}
  \label{eq:me-rot}
  \lefteqn{\dot{\hat{\sigma}} = - 
    \frac{i}{\hbar} [ \hat{H}_{\rm M} \!+\! \hat{H}_{\rm
      R}(t) , \hat{\sigma} ] } & & \\ \nonumber 
  & & + \sum_i  \left[ \hat{S}_i(t) \, \hat{\sigma} \, 
  \hat{S}^\dagger_i(t) - {\textstyle\frac{1}{2}} \, \hat{S}^\dagger_i(t) 
  \hat{S}_i(t) \, \hat{\sigma} - {\textstyle\frac{1}{2}} \, \hat{\sigma}
    \, \hat{S}^\dagger_i(t) \hat{S}_i(t)  \right] ,
\end{eqnarray}
The resulting time dependent Raman Hamiltonian $\hat{H}_{\rm R}(t)$ in
Eq.~(\ref{eq:me-rot}), that describes the stimulated rotational Raman
transitions, reads
\begin{equation}
  \label{eq:raman-ham}
  \hat{H}_{\rm R}(t) = - \frac{1}{\hbar\Delta} \sum_{j_e m_e} 
  \hat{\vec{d}} \!\cdot\! \vec{E}^{(-)}(t) \, | j_e, m_e\rangle 
  \langle j_e, m_e | \, \hat{\vec{d}} \!\cdot\! \vec{E}^{(+)}(t) \; .    
\end{equation}
It describes transitions from rotational states in the electronic
ground state to those in the electronically excited ones and back to
rotational states in the electronic ground state, connected with the
absorption and stimulated emission of laser photons.

Since the dipole operator acts due to its vector character also on the
rotational degrees of freedom, we have to distinguish between
rotational states in the ground and excited electronic states.  Thus
we have denoted this distinction in Eq.~(\ref{eq:raman-ham})
explicitely by the use of the subscript $e$.  If no subscript is used
the electronic ground state is assumed.

The incoherent part of Eq.~(\ref{eq:me-rot}) describes the spontaneous
Raman processes via the time dependent spontaneous Raman operator
\begin{equation}
  \label{eq:def-S}
  \hat{S}_i(t) = \frac{\kappa}{\hbar\Delta}
  \sum_{j_e m_e} 
  \hat{d}_i \, | j_e, m_e \rangle \langle j_e, m_e | \, \hat{\vec{d}}
  \!\cdot\! \vec{E}^{(+)}(t) \; ,
\end{equation}
with $\kappa$ being defined in Eq.~(\ref{eq:def-kappa}). From the
derivation of Eq.~(\ref{eq:me-rot}) it comes out that the Raman
Hamiltonian~(\ref{eq:raman-ham}) and the operator~(\ref{eq:def-S})
are related by
\begin{equation}
  \label{eq:ham-S-relation}
  \sum_i \hat{S}_i^\dagger(t) \hat{S}_i(t) = -
  \frac{\Gamma}{\hbar \Delta}
  \hat{H}_{\rm R}(t) \; .
\end{equation}
This shows that the ratio of spontaneous and stimulated Raman
transitions, inherent in the master equation~(\ref{eq:me-rot}), is
fixed by the ratio $\Gamma/\Delta$ and is independent of the laser
intensity. Thus the laser field must be detuned by several natural
linewidths from the electronic resonance to diminish spontaneous
processes compared to stimulated ones.

A critical issue for the validity of our approach is given by the fact
that we assume a far off-resonant laser field. Clearly spontaneous
Raman scattering will lead to a diffusive broadening of the
vibrational statistics that will slowly populate higher lying
vibrational states. For our calculation to be valid, we do not have to
assume that this process is negligible during the laser pulse
duration. But we have to assume that this process will not populate
higher lying vibrational states, that may eventually be subject to
resonant transitions into an electronically excited state.

For estimating the increase of vibrational excitation during the laser
interaction, a rate equation for the vibrational populations $P_\nu$ can be
derived by omitting the rotational degrees of freedom. It reads
\begin{equation}
  \label{eq:vib-rateeq}
  \dot{P}_\nu = \frac{\Omega_{\rm R} \Gamma}{\Delta} \sum_{\nu'}
  P_{\nu'} \Big( \sum_{\nu_e} | \langle \nu | \nu_e \rangle
  \langle \nu_e | \nu' \rangle |^2 - \delta_{\nu\nu'} \Big) , 
\end{equation}
where the laser-coupling strength is determined by the two-photon Raman Rabi
frequency,
\begin{equation}
  \label{eq:16}
  \Omega_{\rm R} = \frac{|\vec{d} \!\cdot\! \vec{E}^{(+)}|^2}{\hbar^2
  \Delta} .
\end{equation}
The Franck--Condon factors $\langle \nu | \nu_e \rangle$ for transitions
between vibrational state $|\nu\rangle$ and $|\nu_e\rangle$ of the ground and
excited electronic states, respectively, determine the strengths of
spontaneous Raman scatterings between vibrational states in the electronic
ground-state potential.

For vibrationally cold molecules, the internuclear potentials in the ground
and excited electronic states may be approximated by harmonic potentials,
whose centers are shifted by $\Delta x$. Neglecting further the difference of
vibrational frequencies in ground and excited electronic states, the
Franck--Condon factors simplify to overlap integrals of shifted vibrational
eigenstates. Starting from the vibrational ground state, the average
vibrational excitation and variance can then be found from
Eq.~(\ref{eq:vib-rateeq}) as $\bar{\nu}(t) \!=\! g t$ and $\Delta\nu^2(t)
\!=\!  (g t)^2 \!+\! g t ( 1 \!+\!  {\textstyle\frac{3}{4}} \eta^2 )$, where
$g \!=\!  \frac{1}{2} \eta^2 \Omega_{\rm R} \Gamma / \Delta$ and $\eta \!=\!
\Delta x \sqrt{2 \mu \omega_\nu/\hbar}$ is the ratio of $\Delta x$ to the
spatial width of the vibrational ground-state, $\mu$ being the reduced mass.

From the condition $\omega_\nu [\bar{\nu}(t) \!+\! \Delta\nu(t) ] \!\ll\!
\Delta$, which must be fulfilled in order that no resonant single-photon
vibronic transitions occur, the valid dimensionless interaction times $\tau
\!=\! B t$, in units of rotational cycles $B^{-1}$, can be obtained as $\tau
\!\ll\!  [B \Delta^2 / (\Gamma \omega_\nu \Omega_{\rm R})]
\eta^{-2}$. In the weak-field regime, $\Omega_{\rm R} \!<\! B$, and for
large detuning $\Delta$, the first factor is rather large.  Numerical values
of $\eta$ for alkali molecules are from $3.7$ for $\leftexp{7}{\rm Li}_2$ up
to $8.6$ for $\leftexp{133}{\rm Cs}_2$. Thus, also the last factor $\eta^{-2}$
does not severly restrict the range of valid interaction times for our
approximations.


\section{Rotational decoherence in optical Kerr interactions} \label{sec:6} 
As an example for the decoherent effects of spontaneous Raman
processes and as a first application of the master equation derived
here, let us consider a cw monochromatic laser field propagating along
the $z$ direction with linear polarization in $x$ direction. This
configuration implements an optical Kerr interaction, where $\sigma_+$
and $\sigma_-$ polarized field components pick up different phase
shifts depending on the rotational state of the molecule. The field is
then subject to a rotation of its linear polarization while
propagating through the molecular medium. Typically in the treatment
of this optical effect the molecular medium is considered as being in
a stationary (thermal) state, whereas here we focus on the actual
dynamics of the molecule and do not consider the effects on the light
field.

\begin{figure}
  \begin{center}
    \epsfig{file=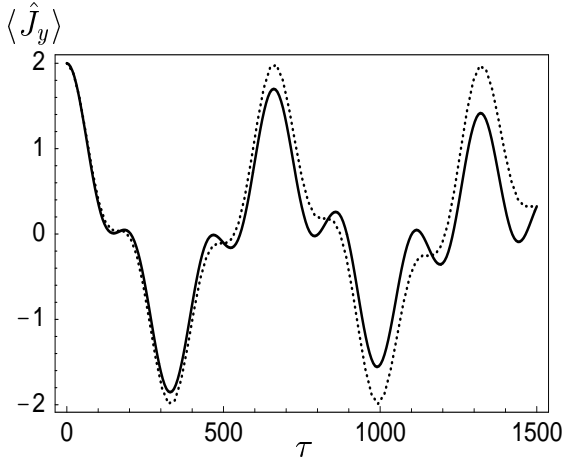, scale=0.92}
    \caption{Dynamics of the average angular-momentum component 
      $\langle \hat{J}_y \rangle$. Initial quantum state is $|j,
      \theta,\phi\rangle$ with $\theta \!=\! \phi \!=\!  \pi / 2$, and
      $j \!=\! 2$, i.e., the initial average angular momentum points
      into $y$ direction. The laser interaction strength is
      $\Omega_{\rm R} / B \!=\! 0.1$ and detuning is $\Delta / \Gamma
      \!=\! 100$. The dimensionless, scaled time is $\tau \!=\! B t$;
      dotted line corresponds to $\Gamma \!=\! 0$.}
    \label{fig:J-y}
  \end{center}
\end{figure} 

Using the quantum-trajectory method, as described in App.~\ref{app:2},
we have numerically calculated the dynamics of the rotational degree
of freedom. We start with an initial rotational quantum state that
corresponds to an average angular momentum pointing in $y$ direction.
It is given by a coherent angular-momentum state, defined as
\begin{eqnarray}
  \label{eq:coh-state}
  | j, \theta, \phi \rangle & = & \sum_{m=-j}^j  {{ 2j \!-\! m \choose m
   }}^{\frac{1}{2}}  [ \cos (\theta/2) ]^{j+m}  [ \sin (\theta
   /2)]^{j-m} \nonumber \\
   & & \qquad \times \, e^{im\phi} |j,m\rangle . 
\end{eqnarray}
We take the quantization axis corresponding to the quantum number $m$
as the $z$ axis. For an average angular momentum pointing in $y$
direction we then choose $\theta \!=\! \pi/2$ and $\phi \!=\! \pi/2$,
which are the spherical angles of the average angular momentum.  The
quantum number $j$ is chosen to be $j \!=\! 2$, as an example for a
rotationally cold molecule that has been weakly excited.
   
The numerically solved temporal evolution of the $y$ component of the
average angular momentum is shown in Fig.~\ref{fig:J-y}.
The laser coupling strength is specified here by the two-photon Raman
Rabi frequency
$\Omega_{\rm R}$, cf.~Eq.~(\ref{eq:16}),
which is chosen in units of the rotational constant $B$ as
$\Omega_{\rm R} / B\!=\! 0.1$. Moreover, the ratio of detuning to
natural linewidth is chosen as $\Delta/\Gamma \!=\! 100$, so that
electronic transitions for a vibrationally cold molecule are widely
suppressed. Nevertheless, rarely occurring spontaneous Raman
transitions are important for understanding the origin of decoherence
effects, as demonstrated for a Raman-driven trapped
ion~\cite{difidio}.

\begin{figure}
  \begin{center}
    \epsfig{file=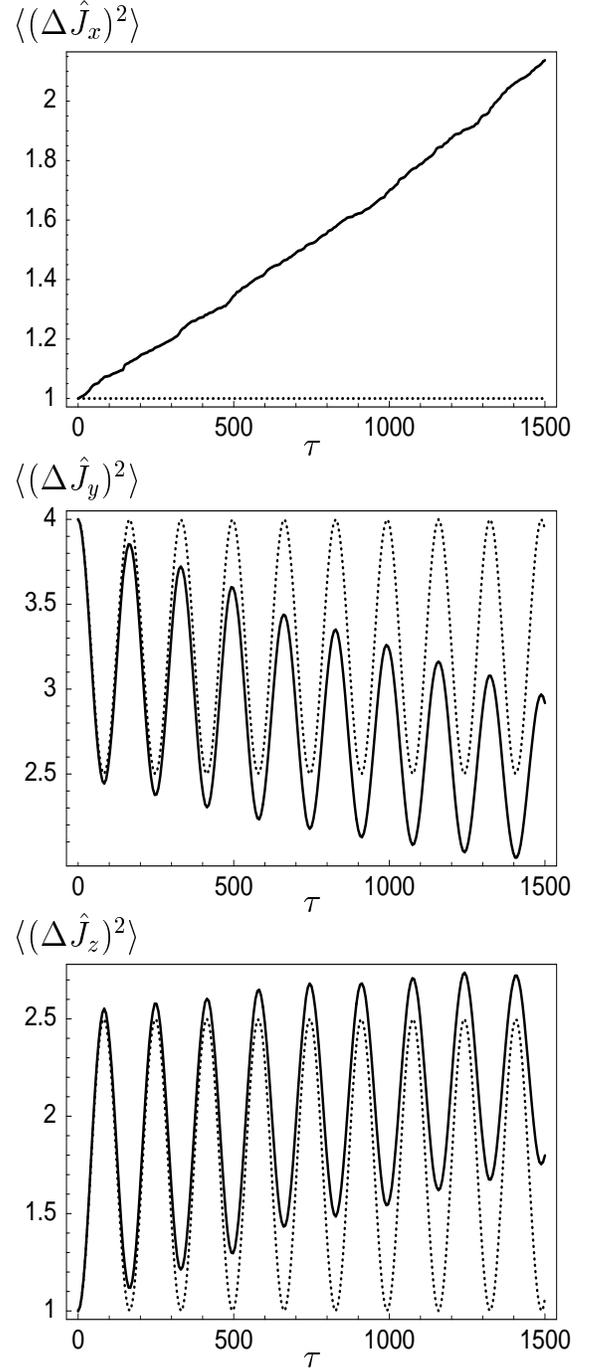,scale=0.92}
    \caption{Time evolution of the fluctuations of the components of
      $\hat{\vec{J}}$ for the same parameters as in Fig.~2 ($\tau
      \!=\! B t$; dotted line corresponds to $\Gamma \!=\! 0$).}
    \label{fig:variances}
  \end{center}
\end{figure}

Neglecting the spontaneous processes ($\Gamma \!=\! 0$) one obtains an
oscillation of the average angular momentum as depicted by the dotted
line in Fig.~\ref{fig:J-y}. In that case the average angular momentum
has no projections on the $x$ and $z$ axes, so that a coherent
one-dimensional oscillation is observed. The angular momentum changes
from positive to negative $y$ projections and vice versa via the $z$
direction. This can be seen in its fluctuations, as shown for the
dotted lines in Fig.~\ref{fig:variances}.
Whereas, the variance in $x$ direction stays at a constant minimal
value, a synchronized exchange of noise is observed in the variances
of the angular-momentum projections in $y$ and $z$ directions.

Taking into account the spontaneous Raman scattering of the laser
light that is detuned by $100$ natural linewidth, coherent processes
are progressively suppressed, cf. solid lines in Figs~\ref{fig:J-y}
and \ref{fig:variances}. The oscillation of the average angular
momentum in $y$ direction is damped, see Fig.~\ref{fig:J-y}. But it
should be noted that no average components in $x$ and $z$ directions
emerge throughout the evolution. Thus the overall coherent dynamics of
the average angular momentum remains a one-dimensional oscillation,
but it is eventually damped to zero.

\begin{figure}
  \begin{center}
    \epsfig{file=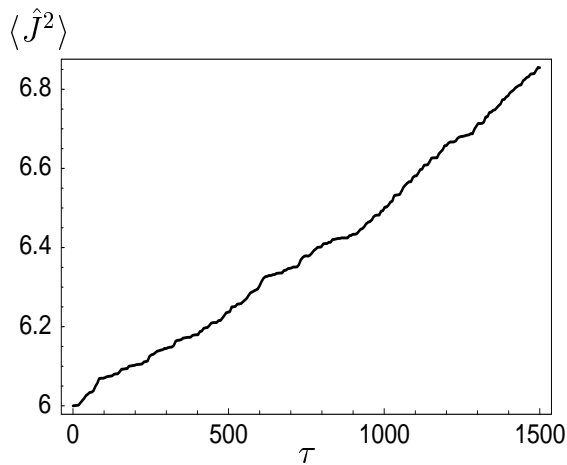,scale=0.92}
    \caption{Heating of the rotational energy by spontaneous Raman
      processes in dependence on the scaled time $\tau \!=\! Bt$;
      parameters are the same as in Fig.~2.}
    \label{fig:L2}
  \end{center}
\end{figure}

On the other hand, fluctuations in overall increase with progressing
time and therefore the average free rotational energy, being
proportional to $\langle \hat{\vec{J}}^2 \rangle$, increases. This
heating is caused by the spontaneous emissions of photons with random
polarizations and at random times. It is given by the sum of all three
variances shown in Fig.~\ref{fig:variances}.  Whereas the oscillation
depth of the noise in $y$ and $z$ directions gradually decreases with
a compensating overall decrease of noise in $y$ and increase of noise
in $z$ directions, there appears a monotonous increase of noise in the
$x$ direction. The latter is due to spontaneous Raman transitions that
excite angular momentum in $x$ direction, but with vanishing mean.
Thus the overall rotational energy is subject to heating and increases
as shown in Fig.~\ref{fig:L2}.

\begin{figure}
  \begin{center}
    \epsfig{file=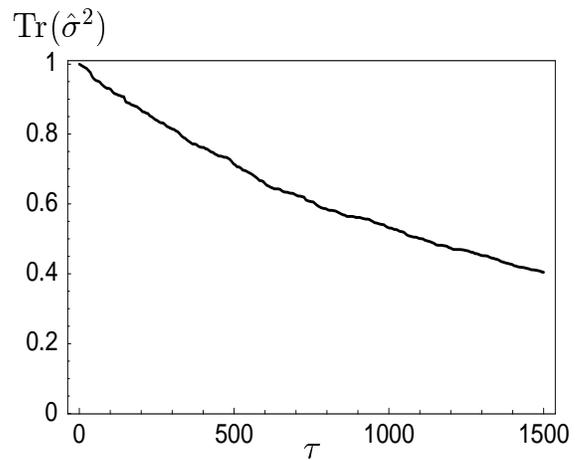,scale=0.92}
    \caption{Purity of the rotational state ${\rm Tr}(\hat{\sigma}^2)$
      over scaled time $\tau \!=\! B t$ for the same parameters as in
      Fig.~2.}
    \label{fig:purity}
  \end{center}
\end{figure}

The decoherence effect of the spontaneous processes can be observed in
the purity of the temporally evolving rotational density operator, see
Fig.~\ref{fig:purity}.  Since any single (experimental) realization of
the time evolution is interrupted by spontaneous emissions of photons
with random polarizations and at random times, the ensemble average,
describing the properties that can be expected from a typical
experiment, gradually develops from an initially pure state into a
statistical mixture.

\section{Summary and conclusions} \label{sec:7}
In summary we have derived a reduced master equation for the rotation
of a small homonuclear molecule for the case of off-resonant laser
excitation with pulse lengths larger than the period of the
internuclear vibration. Several approximations have been justified and
applied for obtaining the sought equations of motion.  First the
electronically excited states have been adiabatically eliminated by
assuming off-resonant laser fields for vibrationally cold molecules.
Second the vibrational degree of freedom has been traced over, based
on the Franck-Condon approximation.

The obtained rotational master equation describes in its coherent part
the interaction of the laser field with the orientation-dependent
polarizability tensor of the molecule. This interaction is generated
by stimulated Raman processes. The incoherent part, on the other hand,
describes the spontaneous Raman processes.  The latter lead to a
damping of the motion of the angular momentum of the molecule and,
more importantly, cause substantial decoherence in the rotational
quantum state.

Our resulting master equation is of Lindblad form and can thus be
efficiently solved by use of quantum trajectories. For the spontaneous
rotational Raman transitions three different quantum jumps can occur,
depending on the polarization of the spontaneously emitted photon. The
importance of these spontaneous processes has been illustrated for the
example of an optical Kerr interaction. Both damping of the rotational
motion and quantum decoherence in the rotational state could be shown.

Our results are relevant for ultracold molecules that are to be
manipulated by off-resonant laser fields. An example is the
interaction of the molecular rotational degree of freedom with the
off-resonant laser fields in a dipole
trap~\cite{pendulum,mol-trap-dipole}. One may expect that even for
trapping purposes decoherence appears in the rotational quantum state.
Considering trapped molecules, our results hold for non-degenerate,
non-interacting molecular gases. One may however guess, that the
decoherence effects derived here will also emerge for degenerate
molecular gases, where a full quantum-field theory applies.  In that
case one may expect a finite lifetime of, for example, molecular
Bose--Einstein condensates due to off-resonant laser interaction with
the molecular rotation, being at the lowest energetic scale.

\begin{acknowledgments}
  The authors thank I.A. Walmsley, E. Tiesinga, P. Domokos, and T.
  Kiss for discussions and comments. This research was supported by
  Deutsche Forschungsgemeinschaft and Deutscher Akademischer
  Austauschdienst.
\end{acknowledgments}

\appendix

\section{Total decay rate} \label{sec:app1}
For obtaining the total decay rate $\Gamma$, we start with the
calculation of
\begin{equation}
  \label{eq:17}
  \Gamma_{ee'} = \sum_i \sum_g \gamma_{i,ge}^\ast \gamma_{i,ge'}
  = \sum_i \sum_g \langle e| \hat{\gamma}_i^\dagger | g \rangle
  \langle g | \hat{\gamma}_i |e' \rangle . 
\end{equation}
Using the definition~(\ref{eq:def-gamma-i}) we obtain
\begin{equation}
  \label{eq:gamma-ee'}
    \Gamma_{ee'} = \kappa^2 \sum_i \sum_g
  \langle e| \hat{d}_i^\dagger |g \rangle \langle g| \hat{d}_i |e'
  \rangle \; .
\end{equation}
For a (homonuclear) molecule having no permanent dipole moment,
$\langle g | \hat{d}_i | g \rangle \!=\! \langle e | \hat{d}_i | e
\rangle \!=\! 0$ and thus the replacement,
\begin{equation}
  \sum_g |g\rangle \langle g | \to \sum_g |g \rangle \langle g | \!+\!
  \sum_e |e \rangle \langle e| = \hat{I}  , 
\end{equation}
can be performed, $\hat{I}$ being the identity operator, to obtain
\begin{equation}
  \label{eq:24}
  \Gamma_{ee'} = \kappa^2 
  \langle e| \hat{\vec{d}} \!\cdot\! \hat{\vec{d}} |e'
  \rangle .
\end{equation}

In Franck--Condon approximation the dipole operator does not depend on
the internuclear coordinate. Moreover, $\hat{\vec{d}} \!\cdot\!
\hat{\vec{d}}$ is a scalar that does not affect the rotational degrees
of freedom. Therefore, we obtain
\begin{eqnarray}
  \label{eq:gamma-ee'2}
    \Gamma_{ee'}  & = & \kappa^2 d^2 \, \delta_{\nu_e \nu_e'} \,
  \delta_{j_e,j_e'} \, \delta_{m_e,m_e'} = \Gamma \delta_{e,e'} ,
\end{eqnarray}
where $d$ is the matrix element of the dipole moment of the considered
electronic transition $|g\rangle \!\leftrightarrow\! |e\rangle$. The
total decay rate $\Gamma$ reads
\begin{equation}
  \label{eq:15}
  \Gamma = \kappa^2 d^2 = \frac{d^2 \omega_{eg}^3}{3 \pi c^3
  \epsilon_0 \hbar} ,
\end{equation}
which is independent of the ro-vibrational state,
$|\nu_e,j_e,m_e\rangle$, in the excited electronic potential.

\section{Quantum trajectories} \label{app:2}
From Eqs~(\ref{eq:raman-ham}) and (\ref{eq:ham-S-relation}) an
effective non-Hermitean Hamiltonian can be defined as
\begin{equation}
  \label{eq:19}
  \hat{H}_{\rm eff}(t) = \hat{H}_{\rm M} + \left( 1 + i
  \frac{\Gamma}{2\Delta}  \right) 
  \hat{H}_{\rm R}(t) \; .
\end{equation}
Using this definition the reduced rotational master
equation~(\ref{eq:me-rot}) can be rewritten as
\begin{equation}
  \label{eq:4}
  \dot{\hat{\sigma}} = - \frac{i}{\hbar} \left[ \hat{H}_{\rm eff}(t)
  \,\hat{\sigma} - \hat{\sigma} \, \hat{H}_{\rm eff}^\dagger(t)
  \right] + \sum_i
  \hat{S}_i(t) \, \hat{\sigma} \hat{S}_i^\dagger(t) \; .
\end{equation}
This equation shows again the interpretation of $\hat{S}_i(t)$ as
being quantum-jump operators that are responsible for spontaneous
Raman processes where a photon is emitted with polarization $i$.

Therefore we may simulate quantum trajectories as realizations of
single experimental runs by sequential application of the non-unitary
evolution operator 
\begin{equation}
  \label{eq:26}
  \hat{U}_{\rm eff}(t,t') = {\cal T} \exp\!\left[ - \frac{i}{\hbar}
  \int_{t'}^t \! d\tau \, \hat{H}_{\rm eff}(\tau) \right] ,
\end{equation}
where ${\cal T}$ denotes time ordering, and the jump operators
$\hat{S}_i(t)$. That is, starting from the normalized state
$|\psi(t_0)\rangle$ at time $t_0$ a quantum trajectory is given by
\begin{eqnarray}
  \label{eq:27}
  \lefteqn{ |\psi(t|\{(t,i)_n,\ldots,(t,i)_1\} \rangle = 
    \hat{U}_{\rm eff}(t,t_n)
    \, \hat{S}_{i_n}(t_n) \ldots} & & \\ \nonumber
   & & \quad \ldots \, \hat{S}_{i_2}(t_2) \, 
   \hat{U}_{\rm eff}(t_2,t_1) \, \hat{S}_{i_1}(t_1)  \, \hat{U}_{\rm
   eff}(t_1,t_0) \, |\psi(t_0)\rangle , 
\end{eqnarray}
where $(t,i)_k$ denote the time $t_k$ of a spontaneous Raman
transition and the polarization $i_k$ of the spontaneously emitted
photon.

Given the trajectory $|\psi(t_n|\{(t,i)_n,\ldots, (t,i)_1\} ) \rangle$
right after a spontaneous Raman process at time $t_n$, the probability
for a further spontaneous process to happen at a later time $t$ is
given by
\begin{widetext}
\begin{eqnarray}
  \label{eq:36}
  P(t) & = & 1 - \frac{\langle  \psi(t|\{(t,i)_n,\ldots,
    (t,i)_1\} )| \psi(t|\{(t,i)_n,\ldots,
    (t,i)_1\} ) \rangle}{\langle
    \psi(t_n|\{(t,i)_n,\ldots, (t,i)_1\} ) |  
    \psi(t_n|\{(t,i)_n,\ldots, (t,i)_1\} ) \rangle}
  \nonumber \\ 
  & = & 1 - \frac{\langle \psi(t_n|\{(t,i)_n,\ldots, (t,i)_1\}) | 
  \exp\!\left[  \frac{\Gamma}{\hbar\Delta}
    \int_{t_n}^t \! d\tau \,
    \hat{H}_{\rm R}(\tau)  \right] |
  \psi(t_n|\{(t,i)_n,\ldots, (t,i)_1\}) \rangle}{\langle
    \psi(t_n|\{(t,i)_n,\ldots, (t,i)_1\} ) |  
    \psi(t_n|\{(t,i)_n,\ldots, (t,i)_1\} ) \rangle} .
\end{eqnarray}
Drawing a random number $R \!\in\! (0,1]$ the condition $P(t) \!=\! R$
determines the time $t \!=\! t_{n+1}$ when the next spontaneous
process occurs. Then to decide which type $i$ of spontaneous Raman
process shall occur, the following probabilities are evaluated
\begin{equation}
  \label{eq:37}
  W_i = \frac{\langle  \psi(t_{n+1}|\{(t,i)_n,\ldots,
    (t,i)_1\} ) | \hat{S}_i^\dagger (t_{n+1}) \hat{S}_i(t_{n+1}) |
    \psi(t_{n+1}|\{(t,i)_n,\ldots, (t,i)_1\} ) 
    \rangle}{\sum_j \langle  \psi(t_{n+1}|\{(t,i)_n,\ldots,
    (t,i)_1\} ) | \hat{S}_j^\dagger (t_{n+1}) \hat{S}_j(t_{n+1}) |
    \psi(t_{n+1}|\{(t,i)_n,\ldots, (t,i)_1\} ) \rangle} \qquad (i=1,2,3) ,
\end{equation}
\end{widetext}
where 
\begin{eqnarray}
  \label{eq:38}
   \lefteqn{ | \psi(t_{n+1}|\{(t,i)_n,\ldots, (t,i)_1\} ) \rangle} && 
   \\ \nonumber 
   & & \quad = \, \hat{U}_{\rm
     eff}(t_{n+1},t_n) \, | \psi(t_n|\{(t,i)_n,\ldots, (t,i)_1\} ) \rangle ,
\end{eqnarray}
and $W_x \!+\! W_y \!+\! W_z \!=\! 1$.  Drawing a second random number
$R' \!\in\! (0,1]$ the type of polarization $i_{n+1}$ of the $(n+1)$th
spontaneously emitted photon is chosen according to
\begin{equation}
  \label{eq:39}
  i = \left\{ \begin{array}{ll}
      x & \mbox{if $0 \!<\! R' \!\leq\! W_x$} \, , \\[0.75ex]
      y & \mbox{if $W_x \!<\! R' \!\leq\! W_x \!+\! W_y$} \, , \\[0.75ex]
      z & \mbox{if $W_x \!+\! W_y \!<\! R' \!\leq\! 1$} \, .  
    \end{array} \right.
\end{equation}
According to the chosen polarization $i_{n+1}$ the resulting
trajectory is then 
\begin{eqnarray}
  \label{eq:40}
   \lefteqn{ | \psi(t_{n+1}|\{(t,i)_{n+1},\ldots, (t,i)_1\} ) \rangle} && 
   \\ \nonumber 
   & & \quad = \, \hat{S}_{i_{n+1}}(t_{n+1}) \, |
   \psi(t_{n+1}|\{(t,i)_n,\ldots, (t,i)_1\} ) \rangle .
\end{eqnarray}


\begin{thebibliography}{99}
  
\bibitem{nonlin-optics} See for example R.W. Boyd, {\it Nonlinear
    optics} (Academic Press, San Diego, 1992); M. Schubert and B.
  Wilhelmi, {\it Nonlinear optics and quantum electronics} (Wiley, New
  York, 1986).
  
\bibitem{qo-book} See for example L. Allen and J.H. Eberly, {\it
    Optical resonance and two-level atoms} (Dover, New York, 1987);
  B.W. Shore, {\it The theory of coherent atomic excitation} (Wiley,
  New York, 1990); W. Vogel, D.-G.  Welsch, and S.  Wallentowitz, {\it
    Quantum optics - an introduction} (Wiley-VCH, Berlin, 2001).

\bibitem{walther-haroche} For the one-atom maser, see
  D. Meschede, H. Walther, and G. M\"uller, Phys. Rev. Lett. {\bf 54},
  551 (1985);
  M. Brune, J.M. Raimond, P. Goy, L. Davidovich, and S. Haroche, {\it
    ibid.} {\bf 59}, 1899 (1987);
%
  G. Rempe, F. Schmidt-Kaler, and H. Walther, {\it ibid.} {\bf 64},
  2783 (1990);
%
  M. Weidinger, B.T.H. Varcoe, R. Heerlein, and H. Walther, {\it
    ibid.} {\bf 82}, 3795 (1999).
  
\bibitem{haroche-qnd} For quantum non-demolition measurements of
  cavity fields and observations of decoherence, cf.
  M. Brune, S. Haroche, V. Lef{\`e}vre, J.M. Raimond, and N. Zagury,
  Phys. Rev. Lett. {\bf 65}, 976 (1990);
  M. Brune, E. Hagley, J. Dreyer, X. Maitre, A. Maali, C. Wunderlich,
  J.M. Raimond, and S. Haroche, {\it ibid.} {\bf 77}, 4887 (1996);
  G. Nogues, A. Rauschenbeutel, S. Osnaghi, M. Brune, J.M. Raimond,
  and S. Haroche, Nature (London) {\bf 400}, 239 (1999).

\bibitem{walther2} For the preparation of cavity photon-number states, cf.
  B.T.H. Varcoe, S. Brattke, M. Weidinger, and H. Walther, Nature
  (London) {\bf 403}, 743 (2000);
  S. Brattke, B.T.H. Varcoe, and H. Walther, Phys. Rev. Lett. {\bf
    86}, 3534 (2001).

  
\bibitem{kimble-rempe} For motional dynamics and trapping of single
  atoms in optical cavity fields, see
  C.J. Hood, M.S. Chapman, T.W. Lynn, and H.J. Kimble, Phys. Rev.
  Lett. {\bf 80}, 4157 (1998);
  P. M\"unstermann, T. Fischer, P. Maunz, P.W.H. Pinkse, and G. Rempe,
  {\it ibid.} {\bf 82}, 3791 (1999);
  J. Ye, D.W. Vernooy, and H.J. Kimble, {\it ibid.} {\bf 83}, 4987
  (1999);
  P.W.H. Pinkse, T. Fischer, P. Maunz, and G. Rempe, Nature (London)
  {\bf 404}, 365 (2000);
  C.J. Hood, T.W. Lynn, A.C. Doherty, A.S. Parkins, and H.J.  Kimble,
  Science {\bf 287}, 1457 (2000);
  P. Horak, H. Ritsch, T. Fischer, P. Maunz, T. Puppe, P.W.H. Pinkse,
  and G. Rempe, Phys. Rev. Lett. {\bf 88}, 043601 (2002);
  T. Fischer, P. Maunz, P.W.H. Pinkse, T. Puppe, and G. Rempe, {\it
    ibid.} {\bf 88}, 163002 (2002).


\bibitem{ion-cooling} For cooling of trapped ions, see D.J. Wineland,
  R. Drullinger, and D.F. Walls, Phys. Rev. Lett. {\bf 40}, 1639
  (1978); W. Neuhauser, M. Hohenstatt, P.E. Toschek, and H. Dehmelt,
  {\it ibid.} {\bf 41}, 233 (1978); F. Diedrich, J.C. Bergquist, W.M.
  Itano, and D.J. Wineland, {\it ibid.} {\bf 62}, 403 (1989); C.
  Monroe, D.M. Meekhof, B.E. King, S.R. Jefferts, W.M. Itano, and D.J.
  Wineland, {\it ibid.} {\bf 75}, 4011 (1995).
  
\bibitem{ion-coupling} For the experimental realization of coherent
  ion-laser interactions, see D.M. Meekhof, C. Monroe, B.E. King, W.M.
  Itano, and D.J. Wineland, Phys. Rev. Lett. {\bf 76}, 1796 (1996).

\bibitem{atom-cooling} For reviews on cooling of atoms, see S. Chu,
  Rev. Mod.  Phys. {\bf 70}, 685 (1998); C.N. Cohen--Tannoudji, {\it
    ibid.} {\bf 70}, 707 (1998); W.D.  Phillips, {\it ibid.} {\bf 70},
  721 (1998), and references therein.
  
\bibitem{alkali-BEC} For first experiments achieving Bose--Einstein
  condensation of alkali atoms, see
  M.H. Anderson, J.R. Ensher, M.R. Matthews, C.E.  Wieman, and E.A.
  Cornell, Science {\bf 269}, 198 (1995);
  %
  C.C. Bradley, C.A. Sackett, J.J.  Tollett, and R.G. Hulet, Phys.
  Rev. Lett. {\bf 75}, 1687 (1995);
  %
  K.B.  Davies, M.-O. Mewes, M.R. Andrews, N.J.  van Druten, D.S.
  Durfee, D.M. Kurn, and W. Ketterle, {\it ibid.\/} {\bf 75}, 3969
  (1995);
  %
  C.C. Bradley, C.A. Sackett, and R.G. Hulet, {\it ibid.\/} {\bf 78},
  985 (1997).
  
\bibitem{hydrogen-BEC} For Bose--Einstein condensation of hydrogen,
  see
  D.G. Fried, T.C. Killian, L. Willmann, D.  Landhuis, S.C. Moss, D.
  Kleppner, and T.J. Greytak, Phys. Rev. Lett. {\bf 81}, 3811 (1998).
  
\bibitem{atom-laser} For implementations of atom lasers, cf. M.-O.
  Mewes, M.R. Andrews, D.M. Kurn, D.S.  Durfee, C.G. Townsend, and W.
  Ketterle, Phys. Rev. Lett. {\bf 78}, 582 (1997);
  %
  I. Bloch, T.W. H\"ansch, and T. Esslinger, {\it ibid.\/} {\bf 82},
  3008 (1999);
  %
  E.W. Hagley, L. Deng, M. Kozuma, J. Wen, K.  Helmerson, S.L.
  Rolston, and W.D. Phillips, Science {\bf 283}, 1706 (1999).
  
\bibitem{atom-microstructure} For experiments with atomic gases in
  microstructures, see J. Reichel, W. H\"ansel, and T.W.  H\"ansch,
  Phys. Rev. Lett. {\bf 83}, 3398 (1999);
  %
  D. Cassettari, B.  Hessmo, R. Folman, T. Maier, and J. Schmiedmayer,
  {\it ibid.\/} {\bf 85}, 5483 (2000);
  %
  W. H\"ansel, J. Reichel, P.  Hommelhoff, and T.W.  H\"ansch, {\it
    ibid.\/} {\bf 86}, 608 (2001); J.  Reichel, W.  H\"ansel, P.
  Hommelhoff, and T.W. H\"ansch, Appl.  Phys.  B {\bf 72}, 81 (2001);
  %
  H. Ott, J. Fortagh, G. Schlotterbeck, A. Grossmann, and
  C. Zimmermann, Phys. Rev. Lett. {\bf 87}, 230401 (2001).

\bibitem{pendulum}
  B. Friedrich and D. Herschbach, Phys. Rev. Lett. {\bf 74}, 4623
  (1995).

\bibitem{mol-trap-dipole} For optically trapping molecules, see
  T. Takekoshi, B.M. Patterson, and R.J. Knize, Phys. Rev. Lett. {\bf
    81}, 5105 (1998).
  
\bibitem{mol-trap-magn} Magnetic trapping of molecules has been
  implemented in
  N. Vanhaecke, W. de Souza Melo, B. Laburthe Tolra, D. Comparat, and
  P.  Pillet, Phys. Rev. Lett. {\bf 89}, 063001 (2002).

  
\bibitem{mol-photoass} For photoassiciation of cold molecules, see for
  example 
  A. Fioretti, D. Comparat, A. Crubellier, O. Dulieu, F.
  Masnou-Seeuws, and P. Pillet, Phys. Rev. Lett. {\bf 80}, 4402
  (1998);
  A.N. Nikolov, E.E. Eyler, X.T. Wang, J. Li, H. Wang, W.C. Stwalley,
  and P.L. Gould, {\it ibid.} {\bf 82}, 703 (1999);
  J.P. Shaffer, W. Chalupcak, and N.P. Bigelow, Phys. Rev. A {\bf 63},
  021401(R) (2001).
  
\bibitem{mol-photoass-bec} For photoassociation starting from an
  atomic condensate, cf.
  R. Wynar, R.S. Freeland, D.J. Han, C. Ryu, and D.J. Heinzen, Science
  {\bf 287}, 1016 (2000);
%
  J.M. Gerton, D. Strekalov, I. Prodan, and R.G. Hulet, Nature
  (London) {\bf 408}, 692 (2000);
%
  C. McKenzie, J. Hecker Denschlag, H. H\"affner, A. Browaeys, L.E.E.
  de Araujo, F.K. Fatemi, K.M. Jones, J.E. Simsarian, D. Cho, A.
  Simoni, E. Tiesinga, P.S. Julienne, K. Helmerson, P.D. Lett, S.L.
  Rolston, and W.D. Phillips, Phys. Rev. Lett. {\bf 88}, 120403
  (2002).


\bibitem{deflection-stapelfeldt} H. Stapelfeldt, H. Sakai, E.
  Constant, and P.B. Corkum, Phys. Rev. Lett. {\bf 79}, 2787 (1997).
  
\bibitem{deflection-sakai} H. Sakai, A. Tarasevitch, J. Danilov, H.
  Stapelfeldt, R.W. Yip, C. Ellert, E. Constant, and P.B. Corkum,
  Phys. Rev. A {\bf 57}, 2794 (1998).

\bibitem{deflection-domokos} P. Domokos, T. Kiss, and J. Janszky, Eur.
  Phys. J. D {\bf 14}, 49 (2001).

\bibitem{brif-decoherence} C. Brif, H. Rabitz, S. Wallentowitz, and
  I.A. Walmsley, Phys. Rev. A {\bf 63}, 063404 (2001).
  
\bibitem{wal-decoherence} S. Wallentowitz, I.A. Walmsley, L.J. Waxer,
  and Th. Richter, J. Phys. B {\bf 35}, 1967 (2002).
  
\bibitem{cohen-tannoudji} For a derivation see for example C.
  Cohen-Tannoudji, J. Dupont-Roc, and G. Grynberg, {\it Atom-Photon
    Interactions: Basic Processes and Applications} (Wiley, 1992).
  
\bibitem{fs-wavepacket} W.S. Warren, H. Rabitz, and M. Dahleh, Science
  {\bf 259}, 1581 (1993);
  %
  B. Kohler, V. Yakovlev, J. Che, J.L. Krause, M. Messina,
  K.R. Wilson, N. Schwentner, R.M. Whitnell, and Y. Yan,
  Phys. Rev. Lett. {\bf 74}, 3360 (1995).

  
\bibitem{difidio} C. Di Fidio, S. Wallentowitz, Z. Kis, and W. Vogel,
  Phys. Rev. A {\bf 60}, R3393 (1999); C. Di Fidio and W. Vogel, Phys.
  Rev. A {\bf 62}, 031802(R) (2000).
  

\end{thebibliography}
\end{document}